\documentclass[aps,prl,reprint,superscriptaddress]{revtex4-2}

\usepackage[utf8]{inputenc}
\usepackage[colorlinks=true, allcolors=blue]{hyperref}
\usepackage[
  forbid-literal-units=true, 
  display-per-mode=symbol, 
  inline-per-mode=symbol, 
  input-ignore=+, 
  exponent-product = \cdot, 
  range-units=single, 
  list-units=repeat, 
  table-alignment-mode=format, 
  table-number-alignment=center,
]{siunitx}
\usepackage{csquotes}
\usepackage{mathtools}
\usepackage{amssymb}
\usepackage{upgreek}
\usepackage{xspace}
\usepackage[capitalise]{cleveref} 
\usepackage[version=4]{mhchem}
\usepackage{orcidlink}

\newcommand{\Sm}{\ce{^{154}Sm}\xspace}
\newcommand{\HIGS}{HI\texorpdfstring{$\upgamma$}{γ}S\xspace}
\newcommand{\Labr}{\ce{LaBr_3}\xspace}
\newcommand{\st}[3]{\ensuremath{{#1}^{#2}_{#3}}}
\newcommand{\jp}[2]{\ensuremath{{#1}^{#2}}}
\newcommand{\g}{\ensuremath{\upgamma}\xspace} 
\newcommand{\D}[1]{\ensuremath{\operatorname{d}\!{#1}}} 

\DeclareSIUnit{\inch}{in.}
\DeclareSIUnit{\barn}{b}
\DeclareSIUnit\elementarycharge{\text{\ensuremath{e}}}

\begin{document}

\title{Gamma decay of the \texorpdfstring{\Sm}{¹⁵⁴Sm} Isovector Giant Dipole Resonance: Smekal-Raman Scattering as a Novel Probe of Nuclear Ground-State Deformation}


\author{J. Kleemann\,\orcidlink{0000-0003-2596-3762}}
\email[]{jkleemann@ikp.tu-darmstadt.de}
\affiliation{Technische Universität Darmstadt, Department of Physics, Institute for Nuclear Physics, 64289 Darmstadt, Germany} 
\author{N. Pietralla\,\orcidlink{0000-0002-4797-3032}}   \affiliation{Technische Universität Darmstadt, Department of Physics, Institute for Nuclear Physics, 64289 Darmstadt, Germany} 
\author{U. Friman-Gayer\,\orcidlink{0000-0003-2590-5052}}   \altaffiliation[Present address: ]{Vysus Group Sweden AB, 214 21 Malmö, Sweden} \affiliation{Department of Physics, Duke University, Durham, North Carolina 27708-0308, USA} \affiliation{Triangle Universities Nuclear Laboratory, Duke University, Durham, North Carolina 27708, USA} 
\author{J. Isaak\,\orcidlink{0000-0002-4735-8320}}   \affiliation{Technische Universität Darmstadt, Department of Physics, Institute for Nuclear Physics, 64289 Darmstadt, Germany} 
\author{O. Papst\,\orcidlink{0000-0002-1037-4183}}   \affiliation{Technische Universität Darmstadt, Department of Physics, Institute for Nuclear Physics, 64289 Darmstadt, Germany} 
\author{K. Prifti\,\orcidlink{0009-0008-0909-3852}}   \affiliation{Technische Universität Darmstadt, Department of Physics, Institute for Nuclear Physics, 64289 Darmstadt, Germany} 
\author{V. Werner\,\orcidlink{0000-0003-4001-0150}}   \affiliation{Technische Universität Darmstadt, Department of Physics, Institute for Nuclear Physics, 64289 Darmstadt, Germany} 
\author{A. D. Ayangeakaa\,\orcidlink{0000-0003-1679-3175}}   \affiliation{Department of Physics and Astronomy, University of North Carolina at Chapel Hill, North Carolina 27599-3255, USA} \affiliation{Triangle Universities Nuclear Laboratory, Duke University, Durham, North Carolina 27708, USA} 
\author{T. Beck\,\orcidlink{0000-0002-5395-9421}}   \altaffiliation[Present address: ]{Facility for Rare Isotope Beams, Michigan State University, East Lansing, Michigan 48824, USA} \affiliation{Technische Universität Darmstadt, Department of Physics, Institute for Nuclear Physics, 64289 Darmstadt, Germany} 
\author{G. Col\`{o}\,\orcidlink{0000-0003-0819-1633}}   \affiliation{Dipartimento di Fisica, Universit\`a degli Studi di Milano and Istituto Nazionale di Fisica Nucleare, Sezione di Milano, 20133 Milano, Italy} 
\author{M. L. Cort\'{e}s\,\orcidlink{0009-0002-7497-6527}}   \affiliation{Technische Universität Darmstadt, Department of Physics, Institute for Nuclear Physics, 64289 Darmstadt, Germany} 
\author{S. W. Finch\,\orcidlink{0000-0003-2178-9402}}   \affiliation{Department of Physics, Duke University, Durham, North Carolina 27708-0308, USA} \affiliation{Triangle Universities Nuclear Laboratory, Duke University, Durham, North Carolina 27708, USA} 
\author{M. Fulghieri}   \affiliation{Department of Physics and Astronomy, University of North Carolina at Chapel Hill, North Carolina 27599-3255, USA} \affiliation{Triangle Universities Nuclear Laboratory, Duke University, Durham, North Carolina 27708, USA} 
\author{D. Gribble\,\orcidlink{0000-0003-4458-3271}}   \affiliation{Department of Physics and Astronomy, University of North Carolina at Chapel Hill, North Carolina 27599-3255, USA} \affiliation{Triangle Universities Nuclear Laboratory, Duke University, Durham, North Carolina 27708, USA} 
\author{K. E. Ide\,\orcidlink{0000-0003-2405-329X}}   \affiliation{Technische Universität Darmstadt, Department of Physics, Institute for Nuclear Physics, 64289 Darmstadt, Germany} 
\author{X. K.-H. James\,\orcidlink{0009-0000-9959-2373}}   \affiliation{Department of Physics and Astronomy, University of North Carolina at Chapel Hill, North Carolina 27599-3255, USA} \affiliation{Triangle Universities Nuclear Laboratory, Duke University, Durham, North Carolina 27708, USA} 
\author{R. V. F. Janssens\,\orcidlink{0000-0001-7095-1715}}   \affiliation{Department of Physics and Astronomy, University of North Carolina at Chapel Hill, North Carolina 27599-3255, USA} \affiliation{Triangle Universities Nuclear Laboratory, Duke University, Durham, North Carolina 27708, USA} 
\author{S. R. Johnson\,\orcidlink{0009-0004-3440-5070}}   \affiliation{Department of Physics and Astronomy, University of North Carolina at Chapel Hill, North Carolina 27599-3255, USA} \affiliation{Triangle Universities Nuclear Laboratory, Duke University, Durham, North Carolina 27708, USA} 
\author{P. Koseoglou\,\orcidlink{0000-0003-4520-4448}}   \affiliation{Technische Universität Darmstadt, Department of Physics, Institute for Nuclear Physics, 64289 Darmstadt, Germany} 
\author{Krishichayan\,\orcidlink{0000-0002-1624-6270}}   \affiliation{Department of Physics, Duke University, Durham, North Carolina 27708-0308, USA} \affiliation{Triangle Universities Nuclear Laboratory, Duke University, Durham, North Carolina 27708, USA} 
\author{D. Savran\,\orcidlink{0000-0002-2685-5600}}   \affiliation{GSI Helmholtzzentrum für Schwerionenforschung GmbH, 64291 Darmstadt, Germany} 
\author{W. Tornow\,\orcidlink{0000-0003-4031-6926}}   \affiliation{Department of Physics, Duke University, Durham, North Carolina 27708-0308, USA} \affiliation{Triangle Universities Nuclear Laboratory, Duke University, Durham, North Carolina 27708, USA} 

\date{\today}

\begin{abstract}
  Gamma decays of the isovector giant dipole resonance (IVGDR) of the deformed nucleus \Sm were measured using \st2+1-Smekal-Raman and elastic scattering of linearly polarized, quasimonochromatic photon beams.
  The two scattering processes were disentangled through their distinct angular distributions.
  Their branching ratio and cross sections were determined at six excitation energies covering the \Sm IVGDR.
  Both agree with the predictions of the geometrical model for the IVGDR and confirm \g decay as an observable sensitive to the structure of the resonance.
  Consequently, the data place strong constraints on the nuclear shape, including the degree of triaxiality.
  The derived \Sm shape parameters $\beta=\num{0.2925 \pm 0.0025}$ and $\gamma=\ang{5.0 \pm 1.5}$ agree well with other measurements and recent Monte Carlo Shell-Model calculations.
\end{abstract}

\maketitle

The nuclear isovector giant dipole resonance (IVGDR) is a fundamental mode of excitation inherent to all nuclei~\cite{BermanFultz1975-GDR-CS-Measurements-Review-And-150Nd-Splitting-Fig,EXFOR}.
It dominates their photonuclear responses as it exhausts most~\cite{BermanFultz1975-GDR-CS-Measurements-Review-And-150Nd-Splitting-Fig,HarakehBook} of the appropriate sum rule for the electric dipole strength, i.e. the Thomas-Reiche-Kuhn sum rule~\cite{Kuhn1925-TRK-Sum-Rule,ReicheThomas1925-TRK-Sum-Rule} multiplied by $(1 + \kappa)$, where $\kappa$ is the enhancement factor~\cite{Speth1981-GDR-Review}.
Since its discovery in the early days of nuclear physics~\cite{BotheGentner1937,BaldwinKlaiber1947,BaldwinKlaiber1948}, the IVGDR as a ground-state excitation has continuously attracted a great deal of attention~\cite{BermanFultz1975-GDR-CS-Measurements-Review-And-150Nd-Splitting-Fig,Speth1981-GDR-Review,HarakehBook,Bortignon2019-Book,Goriely2019-PSF-Database}.
It has also been investigated as a function of temperature~\cite{Bracco2019,Wieland2006,Kelly1999,Bracco1995} and excitations analogous to the nuclear IVGDR were identified in other multiparticle systems such as atoms~\cite{Ederer1964-GDR-Analogues-Atoms,Lukirskii1964-GDR-Analogues-Atoms,Samson1966-GDR-Analogues-Atoms,Wendin1973-GDR-Analogues-Atoms,Shiner2011-GDR-Analogues-Atoms,Pabst2013-GDR-Analogues-Atoms}, metallic clusters~\cite{Heer1987-GDR-Analogues-Metallic-Clusters,Hoheisel1988-GDR-Analogues-Metallic-Clusters,Tiggesbaeumker1992-GDR-Analogues-Metallic-Clusters}, and fullerenes~\cite{Gensterblum1991-GDR-Analogues-Fullerenes,Hertel1992-GDR-Analogues-Fullerenes}.
In the geometrical macroscopic liquid-drop model, the nuclear IVGDR is considered to be a collective isovector oscillation of the protons against the neutrons~\cite{GoldhaberTeller1948-GDR-Hydrodynamic-Model,SteinwedelJensen1950-GDR-Hydrodynamic-Model} while, microscopically, it is understood as a collective coherent one-particle-one-hole excitation of the ground state~\cite{Speth1981-GDR-Review,SpethWambach1991-GDR-Theory}.
Its excitation energy ranges between \qtyrange[range-phrase=\ and\ ]{10}{25}{\MeV}, depending on the mass number.
The IVGDR is, hence, particle unbound and decays predominantly by neutron emission, but also with a probability of about \qty{1}{\percent} by internal \g decay~\cite{Beene1990-GDR-208Pb-Gamma-Branching,Boretzky1996-GDR-208Pb-Gamma-Branching}.

As a ground-state excitation, the IVGDR is sensitive to ground-state properties, especially the nuclear shape.
In spherical nuclei its photoabsorption cross section typically has a Lorentzian shape with a width of a few \unit{\MeV}, while in heavy, deformed nuclei a splitting into two overlapping Lorentzians is observed~\cite{BermanFultz1975-GDR-CS-Measurements-Review-And-150Nd-Splitting-Fig}.
This phenomenon is considered to be one of the prime signatures of nuclear deformation and is understood through the axial deformation allowing vibrations of protons against neutrons parallel or perpendicular to the nuclear symmetry axis with differing frequencies~\cite{Danos1958-GDR-Splitting-Model,Okamoto1958-GDR-Splitting-Model}.
The two oscillation modes are commonly assigned $K$ quantum numbers of $K=0$ for the parallel and $K=1$ for the perpendicular vibration, respectively~\cite{HarakehBook}.
This geometrical picture is widely accepted in textbooks~\cite{HarakehBook} due to its appealing simplicity.
The IVGDR's photoabsorption cross section has been studied extensively in photoabsorption and particle-scattering reactions for many nuclei~\cite{BermanFultz1975-GDR-CS-Measurements-Review-And-150Nd-Splitting-Fig,EXFOR}.
At the same time, this macroscopic interpretation has consequences for the \g decay of the IVGDR, especially for the branching ratio of photon scattering reactions to the ground state and to the \st2+1 member of the ground-state rotational band~\cite{Hayward1970-Book-Photonuclear-Reactions,FullerHayward1962-GDR-NRF-Bremsstrahlung,BarNoyMoreh1974-GDR-NRF-ng-Source}.
The latter may be addressed as a Smekal-Raman scattering (SRS) process.
However, the IVGDR's \g decay, a fundamental property associated with its internal structure, was never studied systematically.
Only few measurements exist, which cover either only low excitation energies~\cite{Hass1971-GDR-NRF-ng-Source,BarNoyMoreh1974-GDR-NRF-ng-Source,Jackson1975-GDR-NRF-ng-Source} or lack sufficient resolution and statistics~\cite{FullerHayward1962-GDR-NRF-Bremsstrahlung,NathanMoreh1980-GDR-NRF-Tagger,HoblitNathan1991-GDR-NRF-Tagger}.
Hence, a close experimental assessment of the geometrical model of the IVGDR in deformed nuclei in terms of its prediction of the \g-decay behavior was never performed thus far.

It is the purpose of this letter to report on the first systematic, energy-resolved study of the evolution of the \g-decay behavior of the IVGDR in a deformed nucleus, \Sm used here as an example, and to discuss its implications.

\begin{figure}
  \includegraphics{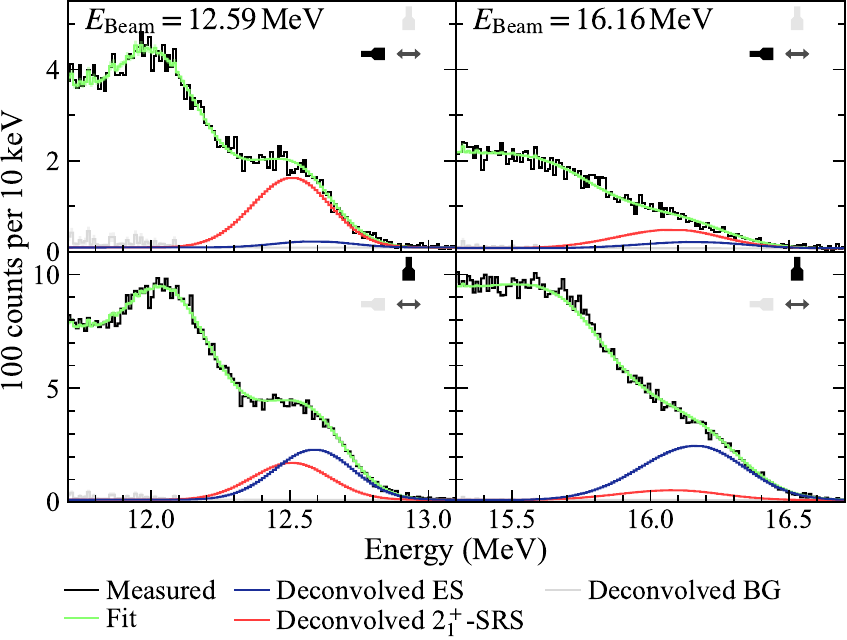}
  \caption{\label{fig:154Sm-GDR-Spectra}%
    Histograms (black) of the \Sm IVGDR \g decay at \qty{12.59}{\MeV} (left) and \qty{16.16}{\MeV} (right) excitation energy measured by \Labr detectors in (top) and perpendicular to (bottom) the beam polarization plane along with the fitted spectra (green).
    The three components (\enquote{Deconvolved}) of the fit before convolution with the detector responses are shown as well, scaled to the full energy peak efficiencies of the detectors, i.e., as responseless detectors would have measured them.
  }
\end{figure}

Nuclear resonance fluorescence (NRF) experiments~\cite{Hayward1970-Book-Photonuclear-Reactions,Zilges2022-NRF-Review,Kneissl1996-NRF-Review,Kneissl2006-NRF-Review,Isaak2022-NRF-Small-Review-2022} on the IVGDR of \Sm have been performed at the High-Intensity $\upgamma$-ray Source (\HIGS)~\cite{Weller-HIGS,Weller2015-HIGS-NuclPhysNews} located at the Triangle Universities Nuclear Laboratory.
By irradiating a \qty{2.42\pm0.08}{\gram\per\centi\meter\squared} \ce{Sm2O3} target enriched to \qty{98.5\pm0.1}{\percent} in \Sm with polarized, quasimonochromatic \g-ray beams provided by \HIGS, the IVGDR of \Sm was photoexcited at different energies. 
With a probability of about \qty{1}{\percent}, the IVGDR promptly decayed internally via the emission of photons.
The latter were detected by four $3\times\qty{3}{\inch}$ \Labr and four high purity Ge clover detectors of the Clover Array setup \cite{Ayangeakaa2021-Clover-Array-Setup} arranged in a close geometry around the target.
The \Labr detectors were mounted at a polar angle of $\theta=\ang{90}$ with respect to the beam direction and azimuthal angles of $\phi\in\{\ang{0}, \ang{90}, \ang{180}, \ang{270}\}$ with respect to the polarization plane, while the clover detectors were placed at $\theta\in\{\ang{125}, \ang{135}\}$ and $\phi\in\{\ang{45}, \ang{180}, \ang{225}, \ang{315}\}$.
Data were taken at mean beam energies, $E_\text{Beam}$, of \qtylist[list-units=single]{11.37; 12.59; 14.27; 15.35; 16.16; 17.79}{\MeV}, chosen to cover the \Sm IVGDR.
The photon beams had slightly asymmetric Gaussian shapes $P(E)$ with bandwidths of $\mathrm{FWHM}/E_\text{Beam}\approx\qty{2}{\percent}$.
The \g signals from elastic scattering (ES) and \st2+1-SRS of the IVGDR both share the shape of the beam profile $P(E)$ since within the narrow energy bandwidths of the beams the cross sections can be considered constant.
In the case of \st2+1-SRS, this peak is additionally shifted down in energy by $E_{\st2+1}=\qty{0.082}{\MeV}$~\cite{NDS154} for \Sm.
Since this shift is smaller than the beam bandwidths, the two signals overlap and cannot be energy resolved in the individual spectra.
However, when exciting the $J^\pi=\jp1-$ IVGDR of even-even nuclei with a linearly polarized photon beam, and assuming a pure electric dipole decay to the \st2+1 state, the angular distributions
\begin{align}
  \label{eq:W_el} W_{\jp0+\to\jp1-\to\jp0+}^{(\text{ES})}(\theta,\phi)        & = \frac{3}{4}(1+\cos^2\theta-\sin^2\theta\cos2\phi)   \\
  \intertext{and}
  \label{eq:W_2}  W_{\jp0+\to\jp1-\to\jp2+}^{(2^+_1\text{-SRS})}(\theta,\phi) & = \frac{3}{40}(13+\cos^2\theta-\sin^2\theta\cos2\phi)
\end{align}
of these scattering processes~\cite{SteffenAlder1975,KraneSteffenWheeler1973,FaggHanna,Iliadis2021} differ significantly at $\theta=\ang{90}$.
Hence, the scattering intensities can be extracted from the observed azimuthal asymmetries of the doublet peaks.
The detector response $\mathcal{R}$, due mostly to escape peaks and the Compton continuum, was taken into account in the analysis, since it dominates the spectra.
Simulations of the responses were performed using \textsc{Geant4}~\cite{Geant4Main,Geant4-06,Geant4-16,UTR2024.01}, taking into account the full experimental setup and the theoretical angular distributions of the observed radiations.
For the analysis of the spectra, first the overall expected radiation was modeled as a sum of two asymmetric Gaussian peaks of the same shape $P(E)$, but different intensities $I$ and shifted by \qty{0.082}{\MeV} with respect to each other on a small background $\mathcal{B}$.
By convolving each part of this model with its corresponding simulated detector response, which includes the angular distribution effects and detector efficiencies, a fit spectrum
\begin{equation}\label{eq:Fit-Spectra}
  \begin{split}
    \mathcal{S}^{d}(E) ={} & I_{\st2+1} \int \mathcal{R}^{d}_{\st2+1}(E^\prime,E)\, P(E^\prime+E_{\st2+1}) \D E^\prime \\
                           & + I_\text{ES} \int\mathcal{R}^{d}_\text{ES}(E^\prime,E)\, P(E^\prime) \D E^\prime         \\
                           & + \int\mathcal{R}^{d}_{\mathcal{B}}(E^\prime,E)\, \mathcal{B}(E^\prime) \D E^\prime
  \end{split}
\end{equation}
is obtained for every detector $d$.
By simultaneously fitting \cref{eq:Fit-Spectra} to all eight observed spectra in a global Bayesian inference Markov chain Monte Carlo approach~\cite{Gelman2014-Bayesian-Data-Analysis,Brooks2011-Bayesian-Data-Analysis,PyMC,PyMC-v5.10.3}, the intensities $I_x$ of ES and \st2+1-SRS on the \Sm IVGDR were extracted from the azimuthal asymmetries and, therefore, also the branching ratio $\sigma_{\st2+1}/\sigma_{\text{ES}}=I_{\st2+1}/I_{\text{ES}}$.
\Cref{fig:154Sm-GDR-Spectra} presents the observed \Sm IVGDR \g-decay spectra of the $\phi\in\{\ang{180}, \ang{90}\}$ \Labr detectors at $E_\text{Beam}\in\{\qty{12.59}{\MeV}, \qty{16.16}{\MeV}\}$ along with the fitted spectra for the determination of the transitions' intensities.
The reliability of the method was confirmed through the analysis of calibration spectra taken with linearly and circularly polarized \g beams on \ce{^{140}Ce} at every beam energy, on \ce{^{12}C} at \qty{15.35}{\MeV}, and on \ce{^{28}Si} at \qty{11.37}{\MeV}.
\ce{^{12}C} and \ce{^{28}Si} each caused one well-defined NRF peak from strong isolated $J^\pi=\jp1+$ resonances~\cite{NDS12,NDS28}, while for \ce{^{140}Ce} the ES on the IVGDR is well isolated, due to the high excitation energy of its first excited state at $E_{\st2+1}=\qty{1.6}{\MeV}$~\cite{NDS140}.
Finally, absolute photon scattering cross sections $\sigma_{\text{ES}}$ and $\sigma_{\st2+1}$ were determined, as well.
This was achieved through simultaneous irradiation of thin \ce{^{197}Au} and \ce{^{nat}CeO2} targets by the \g beam during the experiment and subsequent determination of their activation, due to $(\g,n)$ reactions, which enabled the calibration of the photon fluxes relative to the known $\ce{^{197}Au}(\g,n)\ce{^{196}Au}$ and $\ce{^{140}Ce}(\g,n)\ce{^{139}Ce}$ cross sections~\cite{Fultz1962-GDR-CS-197Au,Lepretre1976-GDR-CS-140Ce}.
\Cref{tab:Primary-Data-Results} provides our primary results.

\begin{table}
  \caption{\label{tab:Primary-Data-Results}%
    Results of the spectra analysis of \Sm.
    $E_\text{Beam}$ denotes the centroid energy of the \g beam with which the IVGDR was photoexcited and FWHM its full width at half maximum.
  }
  \begin{ruledtabular} 
    \begin{tabular}{S[table-format = 2.2(1)] S[table-format = 1.2(1)] S[table-format = 1.4(2)] S[table-format = 1.2(2)] S[table-format = 1.3(2)]
      }
      {$E_\text{Beam}$} & {FWHM}          & {$\sigma_{\st2+1}/\sigma_{\text{ES}}$} & {$\sigma_{\text{ES}}$} & {$\sigma_{\st2+1}$}    \\
      {(\unit{\MeV})}   & {(\unit{\MeV})} & {}                                     & {(\unit{\milli\barn})} & {(\unit{\milli\barn})} \\
      \colrule
      11.37\pm0.01      & 0.33\pm0.01     & 0.978\pm0.023                          & 0.70\pm0.07            & 0.69\pm0.07            \\
      12.59\pm0.01      & 0.27\pm0.03     & 1.017\pm0.015                          & 1.49\pm0.09            & 1.52\pm0.09            \\
      14.27\pm0.01      & 0.28\pm0.01     & 0.878\pm0.017                          & 1.61\pm0.09            & 1.41\pm0.08            \\
      15.35\pm0.01      & 0.34\pm0.01     & 0.436\pm0.034                          & 2.80\pm0.17            & 1.22\pm0.10            \\
      16.16\pm0.01      & 0.32\pm0.01     & 0.255\pm0.0043                         & 3.01\pm0.18            & 0.77\pm0.05            \\
      17.79\pm0.01      & 0.41\pm0.01     & 0.0964\pm0.0032                        & 2.24\pm0.20            & 0.216\pm0.021          \\
    \end{tabular}
  \end{ruledtabular}
\end{table}

\begin{figure}
  \includegraphics{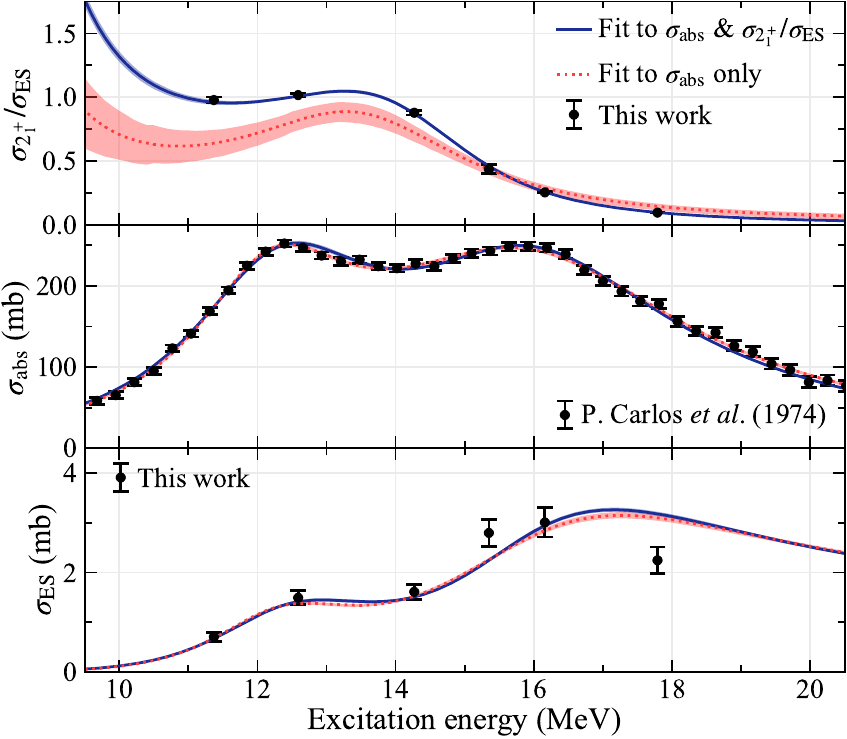}
  \caption{\label{fig:154Sm-GDR-Data-Fits}%
    Simultaneous fit (solid blue) of the SLO parameters of the \Sm IVGDR to both its $\sigma_{\st2+1}/\sigma_{\text{ES}}$ \g-decay behavior (top, data from this work) and its $\sigma_\text{abs}$ photoabsorption cross section (middle, data from Ref.~\cite{Carlos1974-GDR-CS-154Sm}) and fit (dotted red) just to the latter with 1$\sigma$ uncertainty bands.
    The ES cross section (bottom, data from this work) is not explicitly fitted.
  }
\end{figure}

The $\sigma_{\st2+1}/\sigma_{\text{ES}}$ \g-decay branching ratios were interpreted through scattering theory using the geometrical model of the IVGDR.
Since NRF on the IVGDR to the \st0+1 state is indistinguishable from Thomson scattering (ES of photons off the nucleus as a whole), these processes interfere with each other~\cite{Hayward1970-Book-Photonuclear-Reactions,FullerHayward1962-GDR-NRF-Bremsstrahlung,BarNoyMoreh1974-GDR-NRF-ng-Source}.
Hence, the total ES cross section is
\begin{equation}\label{eq:sigma_ES}
  \sigma_{\text{ES}}(E) = \frac{8\pi}{3} \left|f_{\text{IVGDR}}(E) + f_{\text{T}}\right|^2,
\end{equation}
where the $f_x\in\mathbb{C}$ are the forward ($\theta=0$) IVGDR \st0+1 NRF and Thomson scattering amplitudes, respectively.
As both scattering processes share the angular distribution of \cref{eq:W_el}, this was factored out and integrated over the full solid angle.
The Thomson amplitude $f_{\text{T}}=-Z^2e^2/(4\pi \epsilon_0 M c^2)$ for forward scattering on a nucleus of charge $Ze$ and mass $M$ is well-known, real, and energy independent~\cite{Hayward1970-Book-Photonuclear-Reactions},
while the IVGDR \st0+1 NRF amplitude $f_{\text{IVGDR}}(E)$ can be obtained from its total photoabsorption cross section $\sigma_{\text{abs}}(E)$ using the optical theorem and the dispersion relations~\cite{Drechsel2003-Dispersion-Relations} as
\begin{align}
  f_{\text{IVGDR}}(E) & = \frac{E^2}{2\pi^2 \hbar c} \mathcal{P}\!\!\!\!\int_{0}^{\infty} \frac{\sigma_\text{abs}(\epsilon)}{{\epsilon}^2 - E^2} \D \epsilon + \mathrm{i} \frac{E \sigma_\text{abs}(E)}{4\pi \hbar c} \\
                      & = \sum_{k=1}^3 \frac{{\sigma_0}_k E^2 \Gamma_k}{4\pi \hbar c} \frac{\left({E_0}_k^2-E^2\right) + \mathrm{i} E \Gamma_k}{\left({E_0}_k^2-E^2\right)^2+E^2 \Gamma_k^2} \label{eq:f_IVGDR_Sum}   \\
                      & \equiv \sum_{k=1}^3 {f_{\text{IVGDR}}}_k(E)
\end{align}
where $\mathcal{P}\!\!\!\int$ denotes the Cauchy principal value integral and the standard Lorentzian (SLO) parametrization of the IVGDR's photoabsorption cross section
\begin{equation}
  \sigma_\text{abs}(E) = \sum_{k=1}^3 \frac{{\sigma_0}_k E^2 \Gamma_k^2}{\left({E_0}_k^2-E^2\right)^2+E^2 \Gamma_k^2}
\end{equation}
was used in the last step~\cite{Hayward1970-Book-Photonuclear-Reactions,FullerHayward1962-GDR-NRF-Bremsstrahlung,BarNoyMoreh1974-GDR-NRF-ng-Source}.
The sum runs over the three axes of the nuclear \enquote{quadrupoloid} along which the IVGDR can oscillate with the SLO parameters ${E_0}_k$, ${\sigma_0}_k$ and $\Gamma_k$; i.e., the centroid energy, the peak photoabsorption cross section and the FWHM of the respective resonances.
Naturally, the nuclear axes can be degenerate (axially symmetric or spherical nuclei), making the SLO parameters possibly degenerate as well.
For the cross section of SRS~\cite{Bhagavantam1942-Raman-Effect-Book} of the IVGDR to the \st2+1 state,
\begin{equation}\label{eq:sigma_2+1}
  \sigma_{\st2+1}(E) = \frac{4\pi}{3} \sum_{k,l=1}^3 \left| f_{\text{IVGDR}_k}(E) - f_{\text{IVGDR}_l}(E) \right|^2
\end{equation}
is obtained~\cite{MaricMoebius1959-GDR-NRF-CS-Theory,Hayward1970-Book-Photonuclear-Reactions,FullerHayward1962-GDR-NRF-Bremsstrahlung} in the geometrical model with three orthogonal oscillators, where the $f_{\text{IVGDR}_k}(E)$ are the summands in \cref{eq:f_IVGDR_Sum}.
We stress that \cref{eq:sigma_2+1} assumes the \st2+1 state to be the pure rotational excitation of the ground state.
Furthermore, we note that the corresponding Alaga rule~\cite{Alaga1955-Alaga-Rules} for the $\sigma_{\st2+1}/\sigma_{\text{ES}}$ branching ratio is recovered from \cref{eq:sigma_2+1,eq:sigma_ES} when an axially symmetric nucleus with isolated $K=0$ and $K=1$ resonances, i.e. $\Gamma_K \ll |E_{K=1}-E_{K=0}|$, is considered and Thomson scattering is neglected.
To test in detail this modeling of the IVGDR as an isovector oscillation of a quadrupoloid with respect to its three intrinsic major axes, a simultaneous Bayesian inference fit~\cite{Gelman2014-Bayesian-Data-Analysis,Brooks2011-Bayesian-Data-Analysis,PyMC,PyMC-v5.10.3} was performed of the nine SLO parameters to both the new $\sigma_{\st2+1}/\sigma_{\text{ES}}$ \g-decay data and the existing photoneutron data for \Sm~\cite{Carlos1974-GDR-CS-154Sm}.
The latter were corrected for the photon-emission channel to obtain $\sigma_\text{abs}$ photoabsorption data for fitting.
For comparison, a similar fit only on this photoabsorption data was also carried out.
The fits are shown in \cref{fig:154Sm-GDR-Data-Fits}.
While both can reproduce the photoabsorption data, and the ES cross section which was not fitted explicitly, fitting the SLO parameters only to the literature $\sigma_\text{abs}$ photoabsorption cross section gives a nonsatisfactory description of the $\sigma_{\st2+1}/\sigma_{\text{ES}}$ \g-decay data.
A better description is obtained when also fitting to the present data.
This demonstrates the high sensitivity of the \g decay to small differences among the SLO parameters---details that cannot be obtained from the photoabsorption data alone.
Note that this statement also holds when fitting an axially symmetric nuclear shape instead with identical SLO parameters for the second and third IVGDR Lorentzians.
Since the simultaneous fit reproduces both datasets, $\sigma_\text{abs}$ and $\sigma_{\st2+1}/\sigma_{\text{ES}}$, the present data confirm the applicability of the geometrical model of the IVGDR even for its \g decay.
The fitted SLO parameters are given in \cref{tab:SLO-Parameters}.

\begin{table}
  \caption{\label{tab:SLO-Parameters}%
    SLO parameters of the \Sm IVGDR obtained from the simultaneous fit to its $\sigma_{\st2+1}/\sigma_{\text{ES}}$ (this work) and $\sigma_\text{abs}$ (Ref.~\cite{Carlos1974-GDR-CS-154Sm}) datasets.
    See text for details.
  }
  \begin{ruledtabular} 
    \begin{tabular}{l S[table-format = 2.3(2)] S[table-format = 1.3(2)] S[table-format = 3.1(2)]
      }
      {$k$} & {${E_0}_k$ (\unit{\MeV})} & {$\Gamma_k$ (\unit{\MeV})} & {${\sigma_0}_k$ (\unit{\milli\barn})} \\
      \colrule
      1     & 12.384\pm0.030            & 3.305\pm0.084              & 199.3\pm2.5                           \\
      2     & 15.905\pm0.074            & 4.24\pm0.20                & 108.9\pm6.0                           \\
      3     & 16.395\pm0.087            & 5.95\pm0.32                & 103.6\pm5.6                           \\
    \end{tabular}
  \end{ruledtabular}
\end{table}

\begin{figure}
  \includegraphics{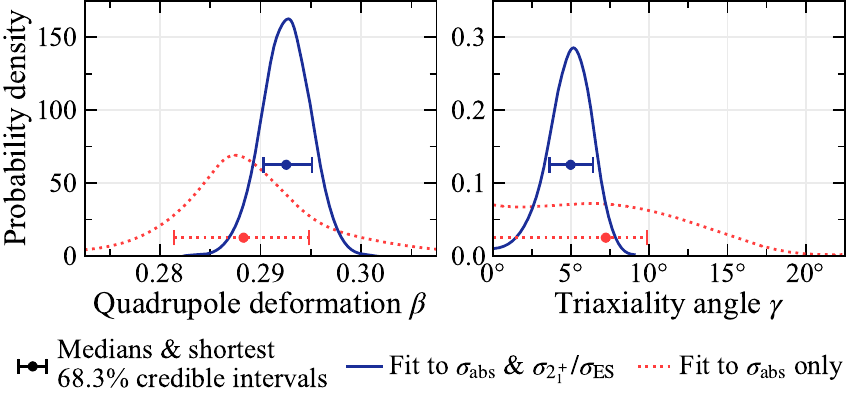}
  \caption{\label{fig:154Sm-GDR-Shape-Fit-Beta-Gamma-Posterior}%
    Posterior distributions for the quadrupole deformation parameter $\beta$ and the triaxiality angle $\gamma$ of \Sm obtained from its fitted IVGDR resonance energies ${E_0}_k$ for both fits shown in \cref{fig:154Sm-GDR-Data-Fits}.
    The uncertainty bars illustrate the medians and shortest $\qty{68.3}{\percent}$ credible intervals of the respective posterior distributions.
  }
\end{figure}

Within the geometrical model, the resonance energy of a IVGDR oscillation is directly related to the length of the nuclear axes $R_k$ along which it takes place~\cite{SteinwedelJensen1950-GDR-Hydrodynamic-Model,Speth1981-GDR-Review}; i.e. 
\begin{equation} \label{eq:E_0_R_Relation}
  {E_0}_k \propto \frac{1}{R_k(\beta, \gamma)} \propto \frac{1}{1 + \sqrt{\frac{5}{4\pi}} \beta \cos\left[\gamma - \frac{2\pi}{3} (k-1)\right]} \text{,}
\end{equation}
where $\beta$ and $\gamma$ are the nuclear deformation parameters~\cite{BohrMot2,DavydovFilippov1958-Triaxiality}.
Hence, the fit of \cref{fig:154Sm-GDR-Data-Fits} also yields information on the nuclear shape of \Sm and, in particular, provides constraints on its nuclear triaxiality from the energy splitting of its second and third IVGDR Lorentzians.
By reparametrizing the ${E_0}_k$ values in the fit through the nuclear deformation parameters according to \cref{eq:E_0_R_Relation}, posterior distributions of the \Sm quadrupole deformation parameter $\beta$ and its triaxiality angle $\gamma$ are obtained through both fits.
These can be found in \cref{fig:154Sm-GDR-Shape-Fit-Beta-Gamma-Posterior}.
Fitting the previously known photoabsorption data only results in a rather flat, featureless posterior distribution for the $\gamma$ deformation with a \qty{99.7}{\percent} credible upper limit of $\ang{28.6}$.
The fit performed while considering in addition the new \g-decay data, however, yields an almost Gaussian distribution centered around \ang{5} with a significantly lower \qty{99.7}{\percent} credible upper limit of $\ang{8.3}$.
Similarly, for the $\beta$ deformation, both posteriors agree overall, but taking into account the \g-decay behavior yields a more constrained distribution with a slightly shifted median.
Taking the median and shortest \qty{68.3}{\percent} credible intervals, $\beta=\num{0.2925 \pm 0.0025}$ and $\gamma=\ang{5.0 \pm 1.5}$ are obtained from the simultaneous fit.
These results match well with the most-recent state-of-the-art configuration interaction calculations in the Monte Carlo Shell Model, which predict $\beta\approx\num{0.28}$ and $\gamma\approx\ang{3.7}$ for the ground-state deformation of \Sm~\cite{OtsukaInPreparation-MCSM-Shape, Otsuka2019-MCSM-Shape}.
The present $\beta$ value disagrees by \qty{16}{\percent} with the value $\beta_2=\num{0.3404\pm0.0017}$ obtained from the $B(E2; \st0+1\to\st2+1)=\qty{4.345\pm0.044}{\elementarycharge\squared\barn\squared}$ transition strength in the compilation of Ref.~\cite{Pritychenko2016-BE2-Tables}.
However, this discrepancy originates from the sole usage of the leading order of the $B(E2)$-to-$\beta^2$ relation and from assuming the solid-angle mean nuclear radius $R_0$ to be $\qty{1.2}{\femto\meter}\times A^{1/3}$ without uncertainty.
Using instead the full relation~\cite{Loebner1970-Deformation-Parameters-Tables,Casten2000}
\begin{equation}
  B(E2; \st0+1 \rightarrow \st2+1) = \left[\frac{3}{4\pi} Z e R_0^2 \left(\beta+\frac{1}{8}\sqrt{\frac{5}{\pi}} \beta^2\right)\right]^2
\end{equation}
and $R_0=\sqrt{5/3}\sqrt{\langle r^2\rangle}$ for the mean nuclear radius~\cite{Krane1987}, with $\sqrt{\langle r^2\rangle}=\qty{5.126\pm0.008}{\femto\meter}$ being the root-mean-square charge radius measured in elastic electron scattering on \Sm~\cite{DeVries1987-Nuclear-Radii-From-Electron-Scattering},
$\beta=\num{0.3067\pm0.0017}$ is obtained from the $B(E2; \st0+1\to\st2+1)$ strength~\cite{Pritychenko2016-BE2-Tables}.
This $\beta$ value agrees with the result of the IVGDR fit within \qty{5}{\percent}.
Similarly~\cite{Loebner1970-Deformation-Parameters-Tables,Casten2000}, $\beta=\num{0.304 \pm 0.006}$ is obtained from the spectroscopic quadrupole moment $Q(\st2+1)=\qty{-1.87\pm0.04}{\barn}$ of the \st2+1 state of \Sm of Ref.~\cite{NDS154}.
This in turn matches with the result of the IVGDR fit.
\Cref{tab:Deformation-Parameters} compares the deformation parameters obtained from the different observables and approaches.
\begin{table}
  \caption{\label{tab:Deformation-Parameters}%
    \Sm deformation parameters $\beta$ and $\gamma$ obtained from different approaches and observables.
    See text for details.
  }
  \begin{ruledtabular} 
    \begin{tabular}{l@{}S[table-format={$\approx$}1.4(2)]@{}S[table-format={$\approx$}1.1(2)]}
      Approach and observables                                                                                                                               & {$\beta$}         & {$\gamma$ (\unit{\degree})} \\
      \colrule
      IVGDR $\sigma_\text{abs}$\,\cite{Carlos1974-GDR-CS-154Sm} and $\sigma_{\!\st2+1}\!/\!\sigma_{\text{ES}}$\,(this work)\!                                & 0.2925 \pm 0.0025 & \num{5.0 \pm 1.5}           \\
      $B(E2; \st0+1\to\st2+1)$~\cite{Pritychenko2016-BE2-Tables} and $\sqrt{\langle R_0^2\rangle}$ \cite{DeVries1987-Nuclear-Radii-From-Electron-Scattering} & 0.3067\pm0.0017   &                             \\
      $B(E2; \st0+1\to\st2+1)$ leading order only~\cite{Pritychenko2016-BE2-Tables}                                                                          & 0.3404\pm0.0017   &                             \\
      $Q(\st2+1)$~\cite{NDS154} and $\sqrt{\langle R_0^2\rangle}$ \cite{DeVries1987-Nuclear-Radii-From-Electron-Scattering}                                  & 0.304 \pm 0.006   &                             \\
      Monte Carlo Shell Model~\cite{OtsukaInPreparation-MCSM-Shape, Otsuka2019-MCSM-Shape}                                                                   & {$\approx$}0.28   & {$\approx$}3.7              \\
    \end{tabular}
  \end{ruledtabular}
\end{table}
The remaining small discrepancy may be due to some of the assumptions in the various approaches not being entirely fulfilled.
For example, \Sm may not be exactly rigid, as inferred from, e.g., the confined $\beta$-soft rotor model~\cite{PietrallaGorbachenko2004-CBS-Model,Moeller2012-154Sm-CBS}, where, due to the finite width of the potential well in $\beta$, the wave functions extracted from $E2$ properties have fluctuations.
The latter result in larger $E2$ matrix elements and, hence, larger effective $\beta$ values than under a rigid rotor assumption.
In addition, due to pairing, the physical ground state might be energetically lower than assumed in the geometrical model of the IVGDR.
This would result in an underestimation of $\beta$ in the IVGDR fit.
It should be emphasized, that the latter would not have a significant effect on the determination of the triaxiality.
Considering the overall good agreement of the $\beta$ deformation parameters obtained from the $B(E2; \st0+1\to\st2+1)$ strength, the $Q(\st2+1)$ moment and the IVGDR \g decay, the high sensitivity of the latter to the nuclear shape is validated.
In particular, a noteworthy sensitivity to the triaxiality degree of freedom is achieved, with a small value of $\gamma=\ang{5.0 \pm 1.5}$ for \Sm.
This finding contributes to the contemporary debate on nuclear triaxial shapes~\cite{DavydovFilippov1958-Triaxiality,Otsuka2019-MCSM-Shape,Grosse2017-GDR-Triaxiality,Grosse2017B-GDR-Triaxiality,Lettmann2017-Triaxiality-Publication,Toh2013-Triaxiality-Publication,Guo2007-Triaxiality-Publication,Ayangeakaa2023-Triaxiality-Publication}.

While this is the first measurement of the evolution of the \st2+1-SRS to ES branching ratio of the IVGDR for a deformed nucleus as a function of energy, it is of interest to investigate this quantity in other nuclei such as \ce{^{166}Er}~\cite{Otsuka2019-MCSM-Shape}, \ce{^{164}Dy}~\cite{OtsukaInPreparation-MCSM-Shape} and \ce{^{188}Os}~\cite{Hayashi1984-188Os-Triaxiality-Prediction}, where triaxialities $\gamma>\ang{7}$ are predicted.
Likewise, the \g-decay behavior of the IVGDRs of \ce{^{152}Sm} and \ce{^{150}Nd} would merit further experiments due to proximity of these nuclei to the spherical-to-deformed shape phase transition~\cite{Iachello2001-X5,CastenZamfir2001-152Sm-X5-Realization,Kruecken2002-150Nd-X5-Realization}, which is expected to be accompanied by an increased softness in $\beta$ deformation.
The IVGDR's photoabsorption data for these nuclei and, therefore, the splitting of their IVGDRs have been disputed recently~\cite{Donaldson2018-GDR-Splitting-Revisited}.
The \g decay of the IVGDR is a yet untapped observable highly sensitive both to the structure of the IVGDR and to the nuclear ground state from which it is photoexcited.
Thus, new measurements would provide further insights in these cases.
Moreover, measurement of the \g decay of the IVGDR in spherical nuclei will constrain state-of-the-art nuclear theory~\cite{Lv2021-GDR-gamma-decay-theory}.

In summary, NRF experiments on the IVGDR of the deformed nucleus \Sm were performed at \HIGS to
study the properties associated with its \g decay.
The measured $\sigma_{\st2+1}/\sigma_{\text{ES}}$ ratios are described well with the macroscopic model of the IVGDR.
This result validates the power of this model and establishes \g decay as a novel observable sensitive to the structure of the IVGDR.
Further investigations with this technique have the potential of providing constraints on nuclear shape parameters, especially on the triaxiality parameter, for nuclei throughout the nuclear chart.
They can also provide further insights in instances where the structure of the IVGDR itself is controversial~\cite{Donaldson2018-GDR-Splitting-Revisited}.

The full data and analysis underlying this letter are openly available at the TUdatalib repository of Technische Universität Darmstadt~\cite{Kleemann2024-154Sm-GDR-TUdatalib-Repository} and further details are available in Ref.~\cite{Kleemann2024-Diss}.

\begin{acknowledgments}
  We thank the \HIGS accelerator group for providing perfect conditions for our experiments and L.~Fortunato, H.-W.~Hammer, M.~N.~Harakeh, T.~Otsuka, H.~A.~Weidenmüller, and A.~Zilges for valuable discussions.
  This work has been funded
  by the German state of Hesse's Ministry of Higher Education, Research and the Arts (HMWK) under grant No. LOEWE/2/11/519/03/04.001(0008)/62,
  by the Deutsche Forschungsgemeinschaft (DFG, German Research Foundation) -- Project-ID 499256822 -- GRK 2891,
  by the German Federal Ministry of Education and Research (BMBF) under grant No. 05P21RDEN9,
  and by the U.S. Department of Energy, Office of Nuclear Physics, under grant Nos. DE-FG02-97ER41041 (UNC) and DE-FG02-97ER41033 (TUNL/Duke).
\end{acknowledgments}

\bibliography{154Sm-GDR-Paper.bib}
\end{document}